\documentclass[10pt,a4paper,leqno,fleqn]{article}
\newcommand\BK{Benjamin Kowarsch}

\usepackage[top=2.4cm, bottom=2.4cm]{geometry} 
\usepackage{enumitem} 
\usepackage{listings} 
\usepackage{xcolor} 
\usepackage{url} 

\usepackage{courier}
\usepackage{lmodern}
\makeatletter
\newcommand{\verbatimfont}[1]{\def\verbatim@font{#1}}
\makeatother

\usepackage{fancyhdr} 
\renewcommand{\emph}[1]{\textbf{\textit{#1}}}

\lstdefinestyle{modula2}{
  language=Modula-2,
  frame=none,
  basicstyle=\fontfamily{pcr}\selectfont\footnotesize,
  keywordstyle=\textbf,
  commentstyle=\italicgray
}

\newcommand\italicgray[1]{\textcolor{gray}{\textit{#1}}}

\usepackage[automake, nonumberlist, nogroupskip, style=altlist]{glossaries}
\makeglossaries

\newglossaryentry{offending facility}{
  name=offending facility, plural=offending facilities,
  description={A language facility that is (a) outdated, harmful or bad habit
  forming in general,\linebreak or (b)~violates any of the design principles
  in section \ref{design-principles}} in particular
}
\newglossaryentry{API}{
  name=API,
  description={Application programming interface}
}
\newglossaryentry{foreign definition module}{
  name=foreign definition module,
  description={A definition module that specifies an interface to a
  \gls{foreign API}}
}
\newglossaryentry{compiler directive}{
  name=compiler directive,
  description={A directive within the source to instruct a language processor
  how to process the input}
}
\newglossaryentry{ETH}{
  name={ETH},
  description={Eidgen\"{o}ssisch Technische Hochschule
  -- Swiss Federal Institute of Technology\footnote
  {Originally founded as {\em Federal Polytechnic School},
   it is still known in English as {\em Z\"urich Polytechnic}.}}
}
\newglossaryentry{foreign API}{
  name=foreign API,
  description={An API implemented in a language other than Modula-2}
}
\usepackage{pifont}

\title{A Proposal for a Revision of ISO Modula-2}
\author{\BK, Modula-2 Software Foundation}
\date{\small{May 2020 (arXiv.org preprint)\footnote
{A discussion paper by the same author proposing a revision of IS 10514-1
 was distributed in 2015 to former WG13 participants. Propositions in this
 paper are based on open peer review feedback from the earlier paper.}}}
\makeatletter
\let\Title\@title
\makeatother
\makeatletter
\let\Author\@author
\makeatother

\begin{document}
\verbatimfont{\small\fontfamily{lmtt}\selectfont}
\maketitle

\begin{abstract}

The Modula-2 language was first specified in \cite{Wirth78} by N.~Wirth at
\Gls{ETH} Z\"{u}rich in 1978 and then revised several times. The last revision
\cite{Wirth88} was published in 1988. The resulting language reports included
ambiguities and lacked a comprehensive standard library.

To resolve the ambiguities and specify a comprehensive standard library an
ISO/IEC working group was formed and commenced work in 1987. A base
standard was then ratified and published as IS 10514-1 in 1996 \cite{ISO96}.
Several conforming compilers have since been developed. At least five remain
available of which at least three are actively maintained and one
has been open sourced.
Meanwhile, various deficiencies of the standard have become apparent but
since its publication, no revision and no maintenance has been carried out.

This paper discusses some of the deficiencies of IS 10514-1 and proposes a
limited revision that could be carried out with moderate effort. The scope of the
paper has been deliberately limited to the core language of the base standard
and therefore excludes the standard library.

\end{abstract}

\section{ISO/IEC Standardisation}

\cite{II2020} describes ISO/IEC standardisation procedures, summarised in
brief below.

\subsection{Organisational Overview}
The International Organisation for Standardisation (ISO) and the International
Electrotechnical Commission (IEC) operate a joint technical committee (JTC1)
to develop, publish and maintain information technology standards. JTC1 is
organised into subcommittees (SC) and working groups (WG). The subcommittee
for programming languages is JTC1/SC22. The working group that developed the
ISO Modula-2 standard was JTC1/SC22/WG13. It was disbanded in 2002.

\subsection{Working Groups and their Lifecycle}
The development and maintenance of ISO/IEC standards is carried out within
working groups by technical experts nominated by national standard bodies
wishing to participate and having membership in the relevant technical committee
or subcommittee.

A working group is established by the relevant committee or subcommittee upon
request by a national standard body with voting membership in that
committee. The request has to be supported by at least four other national standard bodies
with voting membership by declaring their intent to participate in the new working group.

When a working group has completed all its work items, it is either put into stand-by
status for future maintenance or disbanded. Once a working group has been disbanded
it cannot be re-instated. For any maintenance on the standard a new working group
must then be established.

\subsection{Standard Development and Review Cycle}
Once a working group has been formed, it is required to produce a standard
or a revised edition of a standard within no more than 36 months. Once a
standard has been published, it is subject to regular reviews every five years
and may either be reconfirmed or withdrawn. Although JTC1/SC22 tends to
reconfirm standards for which no working group exists to carry out
maintenance, this cannot be taken for granted and such standards could in
principle be withdrawn for lack of experts to carry out a proper review
whenever a regular review is due.

\section{Practical Considerations}

\subsection{Challenges}
The primary challenge with a revision of IS 10514-1 is not the technical work
itself but the establishment of a new ISO/IEC working group, in particular the
recruitment of collaborators to carry out the work. There are three detrimental
factors.

\subsubsection{Negative Perception}
One factor is the common perception that participation in standard bodies is
both time consuming and frustrating. Standardisation work is often characterised
by re-opening of issues that had already been resolved, shifting goals and feature
creep. This was certainly the case with JTC1/SC22/WG13. The root causes were
collaborator turnover and a general lack of project management, in particular the
absence of rigorous definition and guarding of scope.

\subsubsection{Time and Cost of Attending Meetings}
Another factor is the time and cost of attending face-to-face working group
meetings. Such meetings are usually held on a rotational basis and require
recurring international travel and hotel accommodation which participants
have to cover out of their own pockets unless they participate as part of their
employment for and on behalf of a larger corporation.

\subsubsection{Membership Fees Charged by National Standard Bodies}
Yet another factor are the prohibitively high membership fees most national
standard bodies are now charging to individuals willing to participate in
standardisation work. In many countries there are arrangements in place that
permit individuals who are faculty members of universities to participate in their
national standard body without being charged any membership fee. However,
in the realm of programming languages, universities are no longer the primary
actors. For a few mainstream programming languages, the actors are large
corporations. And for the vast majority of programming languages and libraries
the actors are small non-profit groups within the open source software movement.
The majority of these groups and their collaborators cannot afford the membership
fees charged by national standard bodies.

\subsection{Constraints}
In order to recruit the expert collaborators required by \cite{II2020} to form a new
ISO/IEC working group, collaborators must be given assurances that (a)~the effort
they will have to invest is reasonably limited both in terms of total work hours and
elapsed calendar time to completion, (b)~there will be no travel involved and
(c)~there will be no monetary cost to them.

\subsection{Approach}
To meet the aforementioned constraints, we propose the following management
approach:

\renewcommand{\labelenumi}{(\arabic{enumi})}
\begin{enumerate}[leftmargin=!, labelindent=-0.75em, itemindent=0em]
\item the size of the working group shall be kept small,
ideally no more than five or six collaborators
\item it shall be attempted to negotiate fee waivers for collaborators who are not
university staff
\item all deliberations shall be undertaken via email, groupware and
teleconferencing
\item the scope of work shall be agreed upfront, once agreed no further items
shall be added, \\
it shall be strictly limited to 10--12~(target) plus 2--3~(reserve) work items
\item work items shall be chosen to follow a pareto distribution by estimated
work effort, \\
75--80\% shall be lowest to moderate effort while 20--25\% may be high effort
\item any work item on which no consensus can be reached shall be removed
from scope
\item a register of potential future work items shall be maintained for any
potential future revision, \\
items removed from scope due to lack of time or consensus may be added
to this register
\item all work shall be completed and a draft revision shall be presented for
ballot within one year
\end{enumerate}

\section{Methodology}

\cite{Kowarsch18} describes design principles for the maintenance of
classical Modula-2 compilers and how they should be applied, weighted
and prioritised. This paper applies the same considerations to IS 10514-1.
For clarity, the terms of reference from \cite{Kowarsch18} are reproduced
in this section.\footnote
{It should be noted that whilst the same principles are applied, the
conclusions must necessarily be different. Standard revision necessitates
change while compiler maintenance must preserve the original specification.}

\subsection{Design Principles}
\label{design-principles}

The following design principles strongly influenced the proposed revisions in
this paper:

\subsubsection{Single Syntax Principle (SSP)}
\label{SSP}

There should be one and \emph{only one} syntax form to express any given
concept \cite{Dijkstra78}.

\subsubsection{Literate Syntax Principle (LSP)}
\label{LSP}

Syntax should be chosen for readability and comprehensibility by a human
reader \cite{Knuth84}.

\subsubsection{Syntax Consistency Principle (SCP)}
\label{SCP}

Syntax should be consistent. Analogous concepts should be expressed by
analogous syntax.

\subsubsection{Principle of Least Astonishment (POLA)}
\label{POLA}

Of any number of possible syntax forms or semantics, the one likely to cause
the least astonishment for a human reader should be chosen and the
alternatives should be discarded \cite{Geoffrey87}.

\subsubsection{Single Responsibility Principle (SRP)}
\label{SRP}

Units of decomposition, such as modules, classes, procedures and functions
should have a single focus and purpose \cite{Martin09}.

\subsubsection{Principle of Information Hiding (POIH)}
\label{POIH}

Implementation specific details should always be hidden, public access should
be denied \cite{Parnas72}.

\subsubsection{Safety Perimeter Principle (SPP)}
\label{SPP}

Facilities that undermine the safety otherwise safeguarded within the language
should be segregated from other facilities \cite[ch.29]{Wirth88}. Their
use should require an explicit expression of intent by the author and be
syntactically recognisable so as to alert the author, maintainer and reader
of the possible implications. This applies and extends the principle of least
privilege \cite{Saltzer74}.

\subsection{Maintenance Objectives}

The primary objectives for the proposed revisions in this paper are:

\renewcommand{\labelenumi}{(\arabic{enumi})}
\begin{enumerate}[leftmargin=!, labelindent=-0.75em, itemindent=0em]
\item to remove facilities that are harmful, outdated or violate
any of [\ref{design-principles}]
\item to update IS 10514-1 with essential modern facilities
\end{enumerate}

\subsubsection{Weighting and Prioritising Objectives}

\noindent The objectives given above may from case to case conflict with one
another. Which objective should be given preference in the event of a conflict
depends on the following factors:

\begin{enumerate}[itemindent=-0.75em]
\item the severity of the \gls{offending facility}
\item the estimated frequency of use of the \gls{offending facility}
\item the effort to update sources impacted by change or removal of the
\gls{offending facility}
\end{enumerate}

\noindent The greater the severity of an \gls{offending facility}, the stronger
is the case for \emph{change} or \emph{removal}; the lower the estimated
frequency of use and the less the effort to update impacted sources, the
stronger the case for \emph{change} or \emph{removal}. In the event that
the estimated frequency of use is high and the effort to update impacted
sources is significant, \emph{deprecation} may be preferable.\footnote
{In the context of a standard revision, deprecation is a precursor to eventual
removal in a subsequent revision.}

\subsection{Mitigation Methods}

Terms of mitigation methods used in this paper have well defined meanings:

\subsubsection{Warning}

A warning shall be issued for each and every use of the
\gls{offending facility}.

\subsubsection{Change}

The \gls{offending facility} shall be replaced with a proposed alternative.

\subsubsection{Deprecation}

A compiler switch to enable and disable the \gls{offending facility} shall
be provided and it shall be disabled by default. When it is enabled, a
deprecation warning shall be issued for each and every use of the
\gls{offending facility}.

\subsubsection{Removal}

Support for the \gls{offending facility} shall be removed altogether.\\

\par\noindent From an educational perspective, the availability of
\glspl{offending facility} promotes bad habits. \emph{Replacement} or
\emph{removal} is therefore generally preferable to \emph{warning} or
\emph{deprecation}.

\subsubsection{Transformation}

To be used in combination with any of change, deprecation or removal.
A conversion program should be provided that transforms source code that uses
\glspl{offending facility} into semantically equivalent source code that
complies with the proposed revisions in this paper.


\section{Lexis}

\subsection{Lexical Alternatives}

IS 5014-1 specifies lexical alternatives \verb|!|, \verb|@|, \verb|<>|, \verb|&| and
\url{~} as synonyms for \verb!|!, \verb|^|, \verb|#|, \verb|AND| and \verb|NOT|.

\subsubsection{Revision}

Synonym symbols \verb|!|, \verb|@|, \verb|<>|, \verb|&| and \url{~} shall be
\emph{removed} from IS 10514-1.

\subsubsection{Rationale}

The availability of alternative symbols is unnecessary and violates SSP [\ref{SSP}].

Symbols \verb|!| and \verb|@| were not part of the original Modula-2 language,
they were added in IS 10514-1 in the mistaken belief that computer systems
with character sets that lacked a vertical bar and caret, i.e.\ mainframe computers
with 6-bit character sets would still be in use. In reality 6-bit character sets had
already been obsolete for 30 years when IS 10514-1 was published and there
had never been any need for these alternatives in Modula-2.

The inequality operator symbol \verb|#| is preferable to its synonym \verb|<>|
because it resembles the mathematical inequality symbol \footnotesize
\raisebox{0.35ex} {$\neq$} \normalsize and as a single character symbol it 
simplifies lexing. Reserved words \verb|AND| and \verb|NOT| are preferable to
their respective synonyms \verb|&| and \url{~} because of consistency: While
there are synonyms for \verb|AND| and \verb|NOT|, there is none for \verb|OR|.
Although \verb!|! could have been used to denote \verb|OR|, it is already used
to separate case labels and would have introduced ambiguity.

\subsubsection{Backwards Compatibility}

The proposed revision may render existing source code incompatible with the
revised standard. However, affected code may be brought into compliance through
lexical \emph{transformation}, using regular expressions and a filter program such
as \verb|sed| or \verb|awk| which would be virtually effortless.\footnote
{Due to the ease and simplicity of lexical transformation, whenever lexical
transformation is possible, a transitional period has been deemed unnecessary
and thus outright removal is considered preferable over deprecation.}

\subsection{Octal Literals}

IS 10514-1 specifies octal literals suffixed with \verb|B| for numeric values
and \verb|C| for character values.

\subsubsection{Revision}
Octal literals shall be \emph{removed} from IS 10514-1.

\subsubsection{Rationale}

The use of octal numbers in programming languages had already been
outdated for 20 years when IS 5014-1 was published. Moreover, the \verb|B|
and \verb|C| suffixes used to denote octal literals in Modula-2 are also legal
digits within hexadecimal literals. This unnecessarily complicates lexing and
it violates POLA [\ref{POLA}] as it is confusing to human readers of the
source code. 

\subsubsection{Substitution}

The built-in \verb|CHR()| function can be used instead of octal character code
literals. It evaluates constant arguments at compile time and accepts both
decimal and hexadecimal arguments.

\subsubsection{Backwards Compatibility}

The proposed revision will likely render existing source code incompatible with
the revised standard. However, affected code may be brought into compliance
through lexical \emph{transformation}.

\subsection{Set Difference Operator}
\abovedisplayshortskip=0pt

IS 5014-1 specifies the minus symbol as set difference operator.

\subsubsection{Revision}
The set difference operator shall be \emph{changed} to the backslash symbol.

\subsubsection{ISO/IEC 80000-2}
The use of the minus symbol as set difference operator is a violation of
ISO/IEC 80000-2 which explicitly states that the minus symbol
should not be used as set difference operator \cite{II2019}.

\subsubsection{Rationale}
Using the minus symbol for set difference is mathematically incorrect, or at best
ambiguous. The proper mathematical symbol for set difference is a reverse
solidus, resembling a backslash.
\begin{equation}
A \, \setminus \, B = \{ x \in A \mid x \notin B \}
\end{equation}
\begin{equation}
A-B = \{ a-b \mid a \in A \land b \in B \}
\end{equation}
\abovedisplayshortskip=4pt
\belowdisplayshortskip=10pt
\begin{equation}
A \, \setminus \, B \neq A-B
\end{equation}
\abovedisplayshortskip=0pt
\belowdisplayshortskip=0pt

\subsubsection{Backwards Compatibility}

The proposed revision will likely render existing source code incompatible with
the revised standard. However, affected code may be brought into compliance
through lexical \emph{transformation}.

\section{Syntax}

\subsection{Multi-Dimensional Arrays}

IS 10514-1 specifies two alternative syntax forms for multi-dimensional
array type declaration.

\subsubsection{Revision}

The long syntax form for multi-dimensional array declaration shall be
\emph{deprecated}.

\subsubsection{Rationale}

The short syntax form for multi-dimensional array type declaration

\lstset{style=modula2}
\begin{lstlisting}
TYPE Matrix = ARRAY [0 .. Cols], [0 .. Rows] OF REAL;
\end{lstlisting}

\noindent is an abbreviation of and thus equivalent to 
\lstset{style=modula2}
\begin{lstlisting}
TYPE Matrix = ARRAY [0 .. Cols] OF ARRAY [0 .. Rows] OF REAL;
\end{lstlisting}

\par\noindent The availability of alternative syntax forms is unnecessary
and it violates SSP [\ref{SSP}] and SCP [\ref{SCP}]. The long form becomes
increasingly impractical when declaring array types of more than two or three
dimensions. The short form is therefore preferable to the long form.

\subsubsection{Backwards Compatibility}

The proposed revision does not impact the compatibility of legacy sources.

\subsection{Universal Type Conversion Syntax}
\label{UTCS}

IS 10514-1 lacks a universal type conversion syntax.

\subsubsection{Revision}

A universal type conversion operator \verb|::| shall be \emph{added} to
IS 10514-1. Its precedence shall be one level above that of operator
\verb|NOT| and one level below that of parentheses and function calls.

It shall permit safe type conversions (a)~between character types,
(b)~between whole number types, (c)~between real number types and
(d)~between whole and real number types.


\subsubsection{Proposed Syntax}

\begin{verbatim}
typeConvExpr :=
  factor '::' typeIdent ;
\end{verbatim}

\subsubsection{Fallback}

Alternatively -- failing to reach consensus -- a pervasive universal safe type
conversion function \verb|CONV()| analogous to function \verb|CAST()| but
with the above semantics shall be \emph{added} instead.

\subsubsection{Rationale}

In a type safe language with name equivalence such as Modula-2, the
assignment of values between variables of different types is severely restricted
by the type regime. It follows that type conversion becomes an important and
frequent operation. A universal type conversion facility is preferable over type
specific conversion functions as it is consistent and simplifies the core language.
It thereby reduces mental load for authors and readers of source code.

\subsection{Local Modules}

IS 10514-1 permits the lexical nesting of module declarations within the
implementation parts of modules. A module declared within another module
is known as a local module.

\subsubsection{Revision}

Local modules shall be \emph{deprecated}.

\subsubsection{Rationale}
If there is sufficient reason to delegate certain responsibilities of a library
module to a local module, then there is also sufficient reason to delegate
those responsibilities to a separate library module. There is no reason why a
local module should be chosen over a separate library.

A local module within a program or library module unnecessarily increases
the line count of the module and thus reduces its readability and maintainability.
It runs counter to the very rationale of decomposing source
code into separate modules in the first place.

Moreover, a test environment for testing a local module must necessarily be
provided within the hosting module.  This further increases clutter and poses
the question whether to release the module with or without the embedded test
environment. By contrast, a proper library module can be imported into any
number of lexically independent test environments.

\subsubsection{Substitution}
Local modules should be removed from their enclosing module and provided as
proper library modules with their own definition and implementation parts. Such
library modules may then be marked for private use [\ref{PrivMod}], exclusive to
the module they were removed from.

\subsubsection{Backwards Compatibility}

The proposed revision does not impact the compatibility of legacy sources.

\subsection{Private-Use Modules}
\label{PrivMod}

IS 10514-1 lacks a facility to mark a library module for private use.

\subsubsection{Revision}

A \gls{compiler directive} shall be \emph{added} that marks a library module
for private use. A warning shall be issued whenever a module so marked is
imported into any scope other than that of an implementation part of any of
its designated client modules. Implementations may provide a policy setting
or compiler switch to treat these warnings as compilation errors instead.

\subsubsection{Proposed Syntax}

The directive shall be recognised after the module header of a
definition module.

\begin{verbatim}
privModPragma :=
  '<*' 'PRIVATETO' '=' clientModuleList '*>' ;
  
clientModuleList :=
  identList ;
\end{verbatim}

\subsubsection{Rationale}

This facility supports the deprecation of local modules by restoring the
private-use aspect of local modules which is lost when they are removed
from their host modules and converted into library modules. The facility is
also useful to discourage the direct use of lower-level library modules,
especially when their \gls{API} is considered unstable and subject to
frequent changes.

\subsection{Foreign Definition Modules}

IS 10514-1 lacks a facility to mark a definition module as a
\gls{foreign definition module}.

\subsubsection{Revision}

A \gls{compiler directive} to mark a definition module
as a \gls{foreign definition module} shall be \emph{added}.
Implementations that provide means to interface to
\glspl{foreign API} shall implement the directive.

\subsubsection{Proposed Syntax}

The directive shall be recognised after the module header of a
foreign definition module.

\begin{verbatim}
ffiPragma :=
  '<*' 'FFI' '=' '"' foreignAPI '"' '*>' ;
  
foreignAPI :=
  'ASM' | 'C' | 'Fortran' | 'Pascal' | ... ;
\end{verbatim}

\subsubsection{Rationale}

A standardised syntax for marking \glspl{foreign definition module}
is a requirement for source code compatibility across different
implementations.

\section{Pervasives}

\subsection{Large Unsigned Type}

IS 10514-1 lacks a large unsigned type.

\subsubsection{Revision}

A pervasive large unsigned type \texttt{LONGCARD} shall be \emph{added}.
The range of the type shall be large enough to express the highest
address of the addressable memory of the underlying architecture.

\subsubsection{Rationale}

For indices of modern filesystems and databases, a large unsigned type
is an absolute requirement. Many existing Modula-2 implementations
provide a \texttt{LONGCARD} type as a language extension.

\subsection{Unicode Character Type}

IS 10514-1 lacks a dedicated Unicode character type.

\subsubsection{Revision}

A dedicated pervasive Unicode character type \texttt{UNICHAR} shall be
\emph{added}. The type shall be able to represent all Unicode code points
and its internal representation shall use \mbox{UCS-4} encoding.

\subsubsection{Rationale}

Universal international character sets were in their infancy when IS 10514-1 was
developed. Meanwhile, use of the Unicode standard \cite{Unicode2019} has become
ubiquitous. Most programming languages now provide both a conventional 7-bit
or 8-bit character type and an extended 16-bit or 32-bit Unicode character type.

\subsection{Unicode Character Constructor Function}

IS 10514-1 lacks a constructor function for values of type \texttt{UNICHAR}.

\subsubsection{Revision}

A pervasive function \texttt{UCHR()} to return a value of type \texttt{UNICHAR}
for a given code point of type \texttt{LONGCARD} shall be \emph{added}.
This is analogous to existing function \texttt{CHR()} in support of type
\texttt{CHAR}.

\subsubsection{Rationale}

The introduction of type \texttt{UNICHAR} necessitates a constructor function
in support of the new type.

\subsection{Conversion Functions}

IS 10514-1 specifies conversion functions \verb|INT()|, \verb|CARD()|,
\verb|FLOAT()|, \verb|LFLOAT()|, \verb|TRUNC()| and \verb|VAL()|.

\subsubsection{Revision}

Pervasive functions \verb|INT()|, \verb|CARD()|, \verb|FLOAT()|,
\verb|LFLOAT()|, \verb|TRUNC()| and \verb|VAL()| shall be \emph{removed}.

\subsubsection{Rationale}

The universal type conversion facility proposed in
[\ref{UTCS}] replaces type specific conversion functions.

\noindent Moreover, there are serious problems with the aforementioned functions.
Their naming and purpose is inconsistent and confusing, violating
SCP [\ref{SCP}], POLA [\ref{POLA}] and SRP [\ref{SRP}].

While functions \verb|INT()| and \verb|CARD()| are named for their return types,
functions \verb|FLOAT()|, \verb|LFLOAT()|, \verb|TRUNC()| and \verb|VAL()| are not.
Functions \verb|FLOAT()| and \verb|LFLOAT()| are confusingly named since
their return types are \verb|REAL| and \verb|LONGREAL|. IS 10514-1 does not
mandate how the real number types are to be implemented. An implementation
may or may not implement them as floating point numbers. The use of identifiers
that suggest a floating point implementation is therefore incorrect and misleading.

Function \verb|TRUNC()| represents both a conversion function and
mathematical function $trunc(x)$. Its primary purpose is conversion, not truncation.
However, its name indicates truncation, not conversion. The name of a function
whose purpose is conversion should indicate conversion and the function should
ideally perform conversion only.

Conversely, the name of a function whose purpose is truncation should indicate
truncation and the function should strictly perform truncation only. Such a function
would then be more appropriately provided by a math library. It is worthwhile noting
that $trunc(x)$ is a relic of early one's complement hardware, since replaced
by $entier(x)$\footnote
{$entier(x)$ rounds towards $-\infty$ while $trunc(x)$ rounds towards $0$.}
on modern two's complement hardware where truncation is less
efficient and leads to the accumulation of rounding errors \cite{Tucker04}. Thus,
if any pervasive function of this kind is desired, it should be \verb|ENTIER()| but
not \verb|TRUNC()|.

Finally, function \verb|VAL()| has the worst possible naming of all.
Instead of indicating conversion in its name, it indicates that it returns a value.
However, this is is entirely meaningless since all functions return values. The
name carries no information about the function's purpose at all.

\subsubsection{Excluded Functions}

It should be noted that we do not propose to remove functions \verb|CHR()| and
\verb|ORD()| because they are are not conversion functions. Character types do
not represent numeric values and consequently there is no conversion between
character types and numeric types. Instead, \verb|CHR()| is a lookup function that
looks up a character by numeric index. Function \verb|ORD()| performs a reverse lookup.

\subsubsection{Backwards Compatibility}

The proposed revision will likely render existing source code incompatible with
the revised standard. Affected code may be brought into compliance
through syntactical \emph{transformation}.

\section{Semantics}

\subsection{Compatibility of NIL}

\verb|NIL| is not compatible with opaque pointer types and procedure types.

\subsubsection{Revision}

\verb|NIL| shall be compatible with opaque pointer types and procedure types.

\subsubsection{Rationale}

The definition of type specific nil values for every opaque pointer type and
procedure type unnecessarily increases code clutter and mental load.
No type safety is gained by this restriction.

\subsubsection{Backwards Compatibility}

The proposed revision does not impact the compatibility of legacy sources.

\subsection{Casting Constant Values}

Function \verb|CAST()| does not accept constant arguments.

\subsubsection{Revision}

Function \verb|CAST()| shall accept constant arguments.

\subsubsection{Rationale}

The definition of intermediate variables and assignment of constant values
for the sole purpose of casting the values unnecessarily increases code clutter
and mental load. No type safety is gained.

\subsubsection{Backwards Compatibility}

The proposed revision does not impact the compatibility of legacy sources.

\subsection{Pointer Initialisation}

IS 10514-1 does not require pointer variables to be initialised.

\subsubsection{Revision}

All pointer variables shall be initialised to \verb|NIL| and deallocation
should reset them to \verb|NIL|.

\subsubsection{Rationale}
The two primary paradigms of Modula-2 are (a)~program decomposition through
data encapsulation and information hiding, and (b)~reliability through type
safety. Opaque pointer types are the primary instrument through which the
former is achieved. The ability to test whether an opaque pointer type has
been allocated or deallocated is central to achieving the latter.

\subsubsection{Backwards Compatibility}

The proposed mitigation does not impact the compatibility of legacy sources.

\subsection{Write Access to Imported Variables}

IS 10514-1 permits write access to imported variables.

\subsubsection{Revision}

Write access to imported variables shall be \emph{deprecated}. 

\subsubsection{Rationale}

Write access to to imported variables violates POIH [\ref{POIH}]. The
original Modula-2 language report \cite[p.88]{Wirth88} states that
``imported variables should be treated as `read-only' objects''.\footnote
{Unfortunately, Wirth did not implement this guideline in any of his compilers.}

\subsubsection{Backwards Compatibility}

The proposed mitigation does not impact the compatibility of legacy sources.

\subsection{Shifting Non-Bitset Values}

Function \verb|SHIFT()| is restricted to operate on values of \verb|BITSET| types.

\subsubsection{Revision}

Function \verb|SHIFT()| shall operate on values of any 8-, 16-, 32-
and, if available 64-bit type.

\subsubsection{Rationale}

The restriction is entirely impractical because it is most likely to require casting
from the original type to a \verb|BITSET| type before the shift operation and
then casting the result back to the original type afterwards. This unnecessarily
increases clutter and makes the source code difficult to read and maintain.
Let us consider the example of a real world hash function below:

\lstset{style=modula2}
\begin{lstlisting}
PROCEDURE valueForNextChar ( hash : Key; ch : CHAR ) : Key;
BEGIN
    RETURN CAST(Key, ORD(ch)) +
           CAST(Key, SHIFT(CAST(BITSET32, (hash)), 6)) +
           CAST(Key, SHIFT(CAST(BITSET32, (hash)), 16)) - hash;
END valueForNextChar;
\end{lstlisting}

\noindent This function body would be far more readable without the casting:

\lstset{style=modula2}
\begin{lstlisting}
RETURN ORD(ch) + SHIFT(hash, 6) + SHIFT(hash, 16) - hash;
\end{lstlisting}

\noindent The \verb|SHIFT()| function has been placed in pseudo-module \verb|SYSTEM|
for a reason: Facilities provided by \verb|SYSTEM| are by their very definition
unsafe. Any import and use of function \verb|SHIFT()| thereby constitutes a
deliberate decision by the developer to break type safety. No type safety is
gained by restricting the argument types for what is already an unsafe
operation.

\subsubsection{Backwards Compatibility}

The proposed revision does not impact the compatibility of legacy sources.

\subsection{Replacing Variant Records with Extensible Records}

IS 10514-1 specifies records with variant fields, also known as variant records.

\subsubsection{Revision}

Variant record types shall be \emph{removed} from IS 10514-1 and
\emph{replaced} with type safe extensible record types described in \cite{Wirth90}.
Slightly deviating from \cite{Wirth90}, a record type shall only be extensible if its
declaration includes a base type specifier. The declaration of an extensible record
type that does not extend another type shall specify \verb|NIL| as its base type.

Also deviating from \cite{Wirth90}, there shall be no special type guard syntax.
Instead, the \verb|CASE| statement alone shall be used to implement type guards.

\lstset{style=modula2}
\begin{lstlisting}
CASE record OF
| TypeA : (* access to fields unique to TypeA *)
| TypeB : (* access to fields unique to TypeB *)
(* etc *)
\end{lstlisting}

\subsubsection{Proposed Syntax}

\begin{verbatim}
recordType :=
  TYPE Ident '=' RECORD ( '(' baseType ')' )? fieldListSeq END ;
  
baseType :=
  'NIL' | typeIdent ;
\end{verbatim}

\subsubsection{Rationale}

Variant records permit writing a variant field as one variant and then reading
it back as another. This constitutes a hidden type cast and is thus an unsafe
operation. The availability of unsafe operations without prior import from
pseudo-module \verb|SYSTEM| is a violation of SPP [\ref{SPP}].

Furthermore, variant record types are fragile. Whenever another variant is
added to a variant field list, all existing client modules have to be recompiled,
This is known as the fragile base class problem \cite{Mikhajlov1998}.
Variant records are therefore not an adequate facility for software extensibility.

By contrast, extensible record types are type safe and they permit
non-fragile extension.

\subsubsection{Backwards Compatibility}

The proposed revision will likely render existing source code incompatible with
the revised standard. Affected code may be brought into compliance through
syntactical \emph{transformation}.

\printglossary[title=Definitions, toctitle=Definitions]

\bibliographystyle{alpha}
\bibliography{A Proposal for a Revision of ISO Modula-2}
\end{document}